\documentclass[aps,prl,twocolumn,superscriptaddress,letterpaper]{revtex4}
\usepackage{mathrsfs}
\usepackage{amsmath}
\usepackage{bm}
\usepackage{color}
\usepackage{graphicx}
\usepackage{hyperref}

\begin{document}


\title{A New Paradigm for Turbulent Transport Across a Steep Gradient in Toroidal Plasmas}
\author{H. S. Xie}
\email[]{Email: huashengxie@gmail.com} \affiliation{Institute for
Fusion Theory and Simulation, Department of Physics, Zhejiang
University, Hangzhou, 310027, People's Republic of China}
\author{Y. Xiao}
\email[]{Email (Corresponding author): yxiao@zju.edu.cn}
\affiliation{Institute for Fusion Theory and Simulation, Department
of Physics, Zhejiang University, Hangzhou, 310027, People's Republic
of China}
\author{Z. Lin} \affiliation{Department of Physics and
Astronomy, University of California, Irvine, California 92697,
USA}\affiliation{Fusion Simulation Center, School of Physics, Peking
University, Beijing 100871, China}

\date{\today}

\begin{abstract}
First principle gyrokinetic simulation of the edge turbulent
transport in toroidal plasmas finds a reverse trend in the turbulent
transport coefficients under strong gradients. It is found that
there exist both linear and nonlinear critical gradients for the
nonmonotonicity of transport characteristics. The discontinuity of
transport flux slope around the turning gradient shows features of
second order phase transition. Under strong gradient the most
unstable modes are in non-ground eigenstates with unconventional
mode structures, which significantly reduces the effective
correlation length and thus reverse the transport trend. Our results
suggest a completely new mechanism for the low to high confinement
mode transition without invoking shear flow or zonal flow. [accepted
by Phys. Rev. Lett.]
\end{abstract}


\maketitle

With input power increasing, a sudden transport `phase' with
formation of edge transport barrier is found experimentally in
fusion plasmas \cite{Wagner1982}, which is called the high (H)
confinement mode to distinguish from the conventional low (L)
confinement mode, where no steep gradients exist in the plasma
profiles. Transport barriers, also recognized in other systems, such
as in geophysical and atmospheric sciences \cite{Or1987}, can be
universal and important. The H-mode plasmas store twice more energy
than that in the L-mode, thus enabling high fusion gain. The H-mode
is the baseline operation scenario of the International
Thermonuclear Experimental Reactor (ITER) \cite{Ikeda2007}. The L-H
transition involves a discontinuous change of the transport
characters and the underlying mechanism remains elusive
\cite{Wagner2007}.  The transition between the multiple equilibrium
states resembles the continuous (or second order) phase transition
of Landau \cite{Landau1937}, a critical phenomenon widely existing
in nature. An improved understanding of the transition physics is
not only important for fusion plasmas, but leads to a new paradigm
for the nonlinear physics in the laboratory and universe.

Several theories have been proposed for the transport characteristic
change \cite{Biglari1990, Hahm1994, Waltz1994, Li2002} or the sudden
L-H transition \cite{Itoh1993, Kim2003}, where generally shear or
zonal flow is invoked. But none of these theories have been fully
verified by first principle simulation or validated by fusion
experiment. In addition, due to many inherent {\it ad hoc}
assumptions on the kinetic physics, these theories may only be
qualitatively correct. Recently, several fluid models (cf.
\cite{Rogers1998}) have also produced some of the essential features
of the L-H transition, i.e., two transport `phases' are found by
increasing input power. These fluid simulation results may not be
conclusive due to overlook of essential kinetic physics. A fully
kinetic simulation of the L-H transition is still precluded due to
the multiple temporal and spatial scale nature of the problem. The
gyrokinetic simulation is so far still one of the best tools to
study the kinetic physics in the edge, although the validity of
gyrokinetics under strong gradients is still under active research.
When studying low frequency drift wave turbulence in the tokamak
edge, the gyrokinetics \cite{Brizard2007} may still be valid for
$R/L_T\sim100$, where $\rho_i/L_T\sim0.1$, satisfying the
gyrokinetic assumption $\rho_i/L_T\ll1$. In this work, we only
consider electrostatic drift wave turbulence by varying
density/temperature gradients while fixing the density/temperature
in the simulation. Our gyrokinetic simulation \cite{Brizard2007,
Lee1987} using the GTC code \cite{Lin1998,Lin2004} shows that both
the linear and nonlinear physical characteristics change
nonmonototically with a turning point at some critical gradient,
which divides the gradient space into a weak gradient region
(L-mode) and a strong gradient region (H-mode). It is known that
drift wave turbulence can lead to anomalous transport
\cite{Horton1998}. It is also commonly accepted that stronger
gradient leads to higher transport coefficients
\cite{Dimits2000,Scott2006}. Based on large scale global gyrokinetic
simulations using the GTC code, we report here for the first time
that the turbulent transport coefficients in toroidal plasmas can be
reversed under strong gradient, i.e., larger gradient leads to
smaller transport coefficient. The slope of the transport flux also
shows a discontinuous change around the turning gradient, similar to
the second order phase transition of Landau \cite{Landau1937}. The
underlying physics is found to be closely related to the
unconventional mode structure under strong gradients, which
significantly reduces the radial correlation length. These novel
findings may suggest a completely new mechanism for the L-H
transition without invoking shear flow or zonal flow.

The GTC code is a well-benchmarked global gyrokinetic particle
simulation code \cite{Lin2004, Rewoldt2007, Xiao2009, Holod2013},
suitable for simulating both electrostatic and electromagnetic drift
wave turbulence \cite{Horton1998}. In the low beta limit, we only
use the electrostatic capability of the GTC code. The simulation
parameters are taken from typical H-mode experiments of the HL-2A
tokamak \cite{Kong2017,Xie2015} using deuterium as the ion species,
with on-axis toroidal magnetic field $B_0=1.35T$, minor radius
$a=40cm$, major radius $R_0=165cm$, safety factor $q=2.7$, magnetic
shear $s=0.5$, plasma temperature $T_e=T_i=200eV$, plasma density
$n_e(r)=n_i= 4.0\times10^{12}cm^{-3}$. Assuming that the time scale
for electron-ion energy exchange is shorter than the profile
relaxation time scale, we set in the simulation the plasma profile
gradients $R_0/L_{T_i}=R_0/L_{T_e}=R_0/L_n$, where $L_{T_i}$,
$L_{T_e}$ and $L_n$ are the scale lengths for ion temperature,
electron temperature and particle density, respectively, i.e.,
$L_T^{-1}\equiv-d\ln T/dr$. Therefore, we keep $\eta=L_n/L_T=1$
throughout this article. We note that one of the most important
parameters is the peaking gradient \cite{Xie2015}. In addition,
circular cross-section is assumed for equilibrium magnetic flux
surface.

\begin{figure}
\centering
\includegraphics[width=8.5cm]{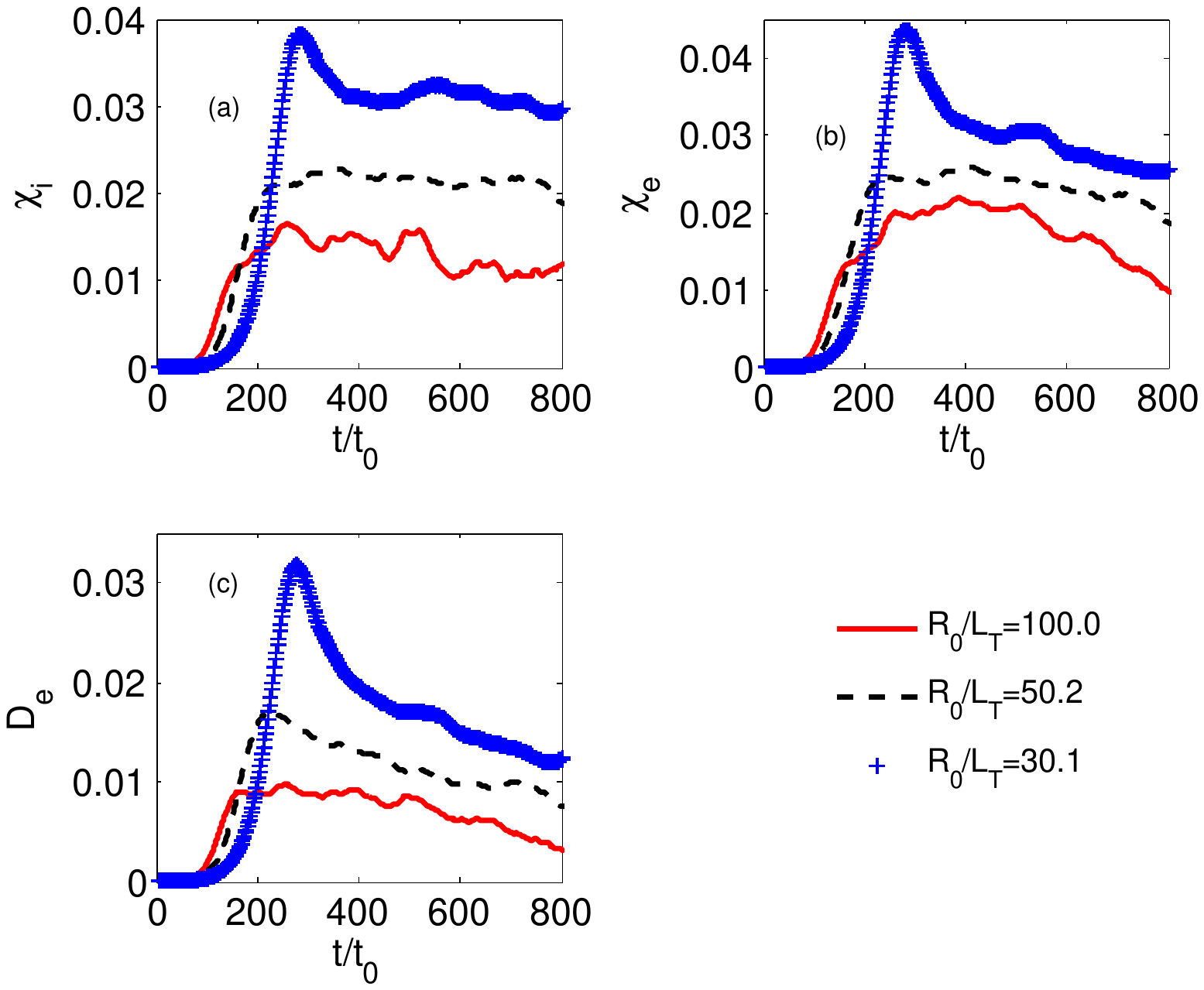}\\
\caption{Time history for three volume averaged physical quantities:
(a) ion heat conductivity; (b) electron heat conductivity; (c)
electron particle diffusivity under three strong temperature
gradients, where $t_0=0.002c_s/R_0$ is time step
size.}\label{fig:hsty_cmp_scangrad_h}
\end{figure}

\begin{figure}
\centering
\includegraphics[width=8.0cm]{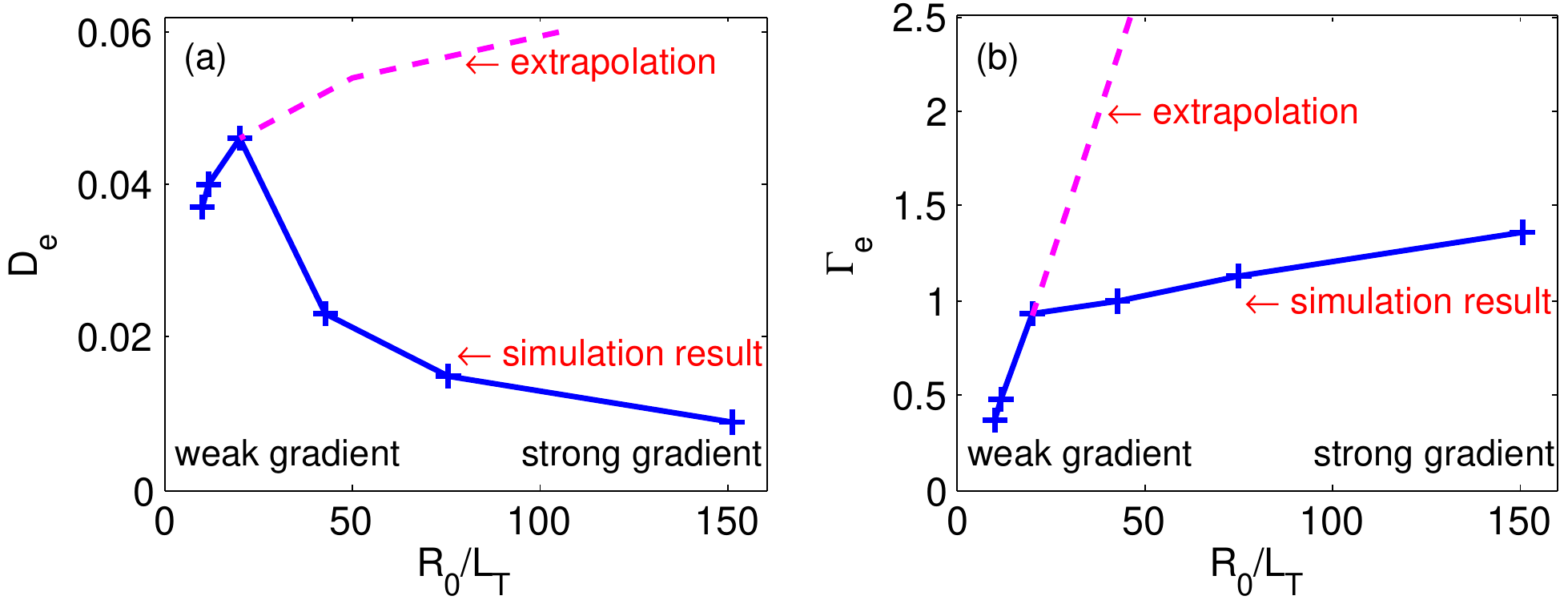}\\
\caption{Time averaged electron particle diffusivity (a) and
electron particle flux (b) for different temperature gradients. A
turning point (critical gradient) is found for the trend of the
transport coefficients.}\label{fig:flux_vs_grad}
\end{figure}

Using the preceding experimental parameters, we carry out a series
of turbulence simulations by scanning the plasma profile gradients.
In the simulations, we use number of grids $150\times1200\times32$
in the radial, poloidal and parallel direction respectively which
leads to a grid size $\sim0.5\rho_i$, and 50 ions/electrons per cell
to reduce the numeric noise. A larger number of grids and more
particles per cell are used and a satisfactory convergence can be
obtained for the simulation results. Zero boundary conditions are
used at $r=0.7a$, and $1.0a$. The time history of the volume
averaged turbulent heat conductivity and particle diffusivity are
shown in Fig. \ref{fig:hsty_cmp_scangrad_h}(a)-(c) for three strong
gradients, $R_0/L_T=30, 50, 100$,  where the heat conductivity
$\chi_j$ is defined from the heat flux $q_j=\int
dv^3(\frac{1}{2}m_jv^2-\frac{3}{2}T_j)\delta v_E\delta f_j\equiv
n_j\chi_j\nabla T_j$, $j=i,e$, and the particle diffusivity $D_j$ is
defined from the particle flux $D_j=\int dv^3\delta v_E\delta f_j$,
with $v_E$ is the $E\times B$ drift caused by turbulence. As shown
in Fig. \ref{fig:hsty_cmp_scangrad_h}, both heat conductivity and
particle diffusivity decrease with the temperature gradient in the
strong gradient region. This is contradictory to the common
knowledge that stronger gradient leads higher transport coefficient
\cite{Dimits2000}. This phenomenon is further illustrated in Fig.
\ref{fig:flux_vs_grad}(a) by comparing the electron diffusivity for
different gradients, where the diffusivity is obtained by the time
average of the saturated value in the nonlinear stage before the
quasilinear flattening of plasma profile occurs. The electron
diffusivity first increases with the gradient, which is consistent
with previous studies \cite{Dimits2000}. However, when we continue
to increase the gradient after some critical value, as shown in Fig.
\ref{fig:hsty_cmp_scangrad_h}, stronger gradient leads to lower
particle diffusivity. The pink dash line is following the
conventional trend by artificial extrapolation \cite{Dimits2000}.
This reverse trend of the transport coefficients also holds for ion
and electron heat transport, which can be seen in Fig.
\ref{fig:hsty_cmp_scangrad_h} (a) or (b). We note that this is the
first time that such extraordinary behavior is observed for the
turbulent transport under strong gradient. In Fig.
\ref{fig:flux_vs_grad}(b) for the electron flux vs. $R_0/L_T$, a
turning point for the gradient drive appears, and the particle flux
reaches a saturation level, or increases much slower, when the
gradient is beyond the turning point, clearly showing a
discontinuous change of the slope of the particle flux. This
provides strong evidence for the formation of gradient transport
barrier, even though no bifurcation occurs. If one adds more power
to the core plasma than the pedestal can exhaust, the L-H transition
can occur due to the gradient transport barrier, which could explain
the mystery of the L-H transition, namely the input power should
exceed a certain level in order to trigger the L-H transition.

\begin{figure}
\centering
\includegraphics[width=8.5cm]{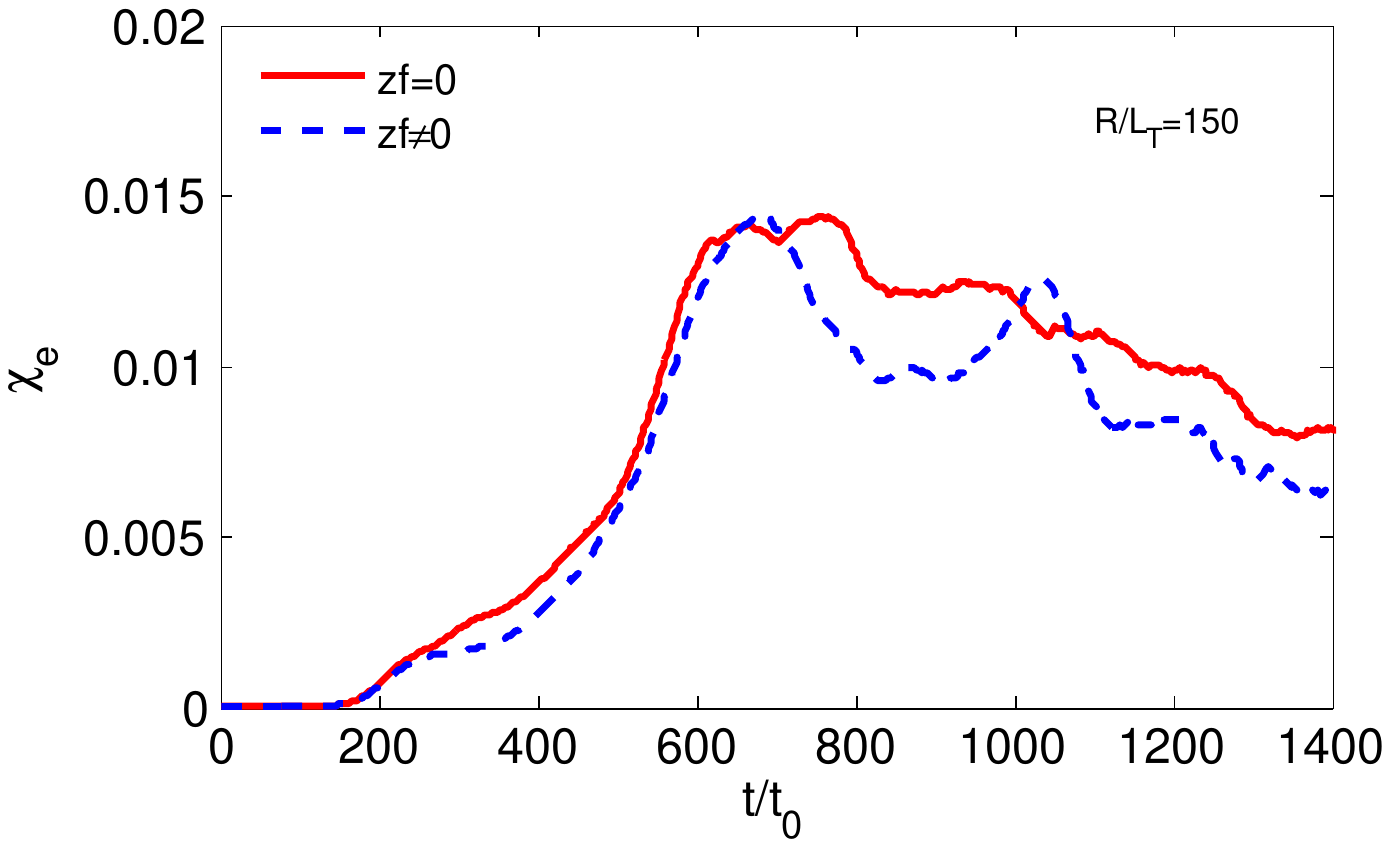}\\
\caption{Time history of electron heat conductivity in the nonlinear
gyrokinetic simulation are shown for two cases under strong
gradient: with and without zonal flow. The green dashed line
represents simulation with zonal flow self-consistently generated.
For the red solid line, the zonal flow is artificially removed from
simulation. }\label{fig:hsty_cmp_zf_lh}
\end{figure}

Next we examine the zonal flow effect on the nonlinear physics under
strong gradient. A previous study using the GTC code has shown that
for the TEM mode the zonal flow can reduce the turbulent transport
significantly ($>50\%$) under weak gradient ($R_0/L_T=6.9$)
\cite{Xiao2009}. However, under strong gradient, Fig.
\ref{fig:hsty_cmp_zf_lh} shows that the zonal flow has little effect
in regulating turbulence. The weak importance of zonal flow near the
edge is also reported in a recent H-mode experiment by Ref.
\cite{Kobayashi2013}.

\begin{figure}
\centering
\includegraphics[width=9.0cm]{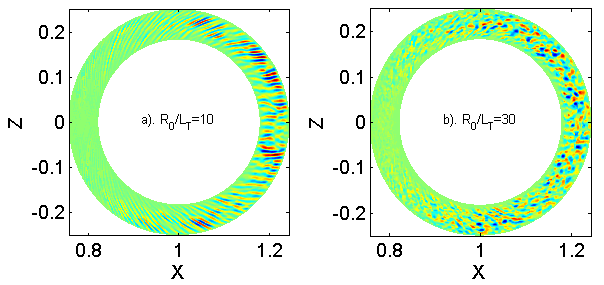}\\
\caption{2D turbulence intensity in the poloidal plane for
nonlinearly weak gradient $R_0/L_T=10$ and nonlinearly strong
gradient $R_0/L_T=30$. }\label{fig:phirz}
\end{figure}

\begin{figure}
\centering
\includegraphics[width=8.5cm]{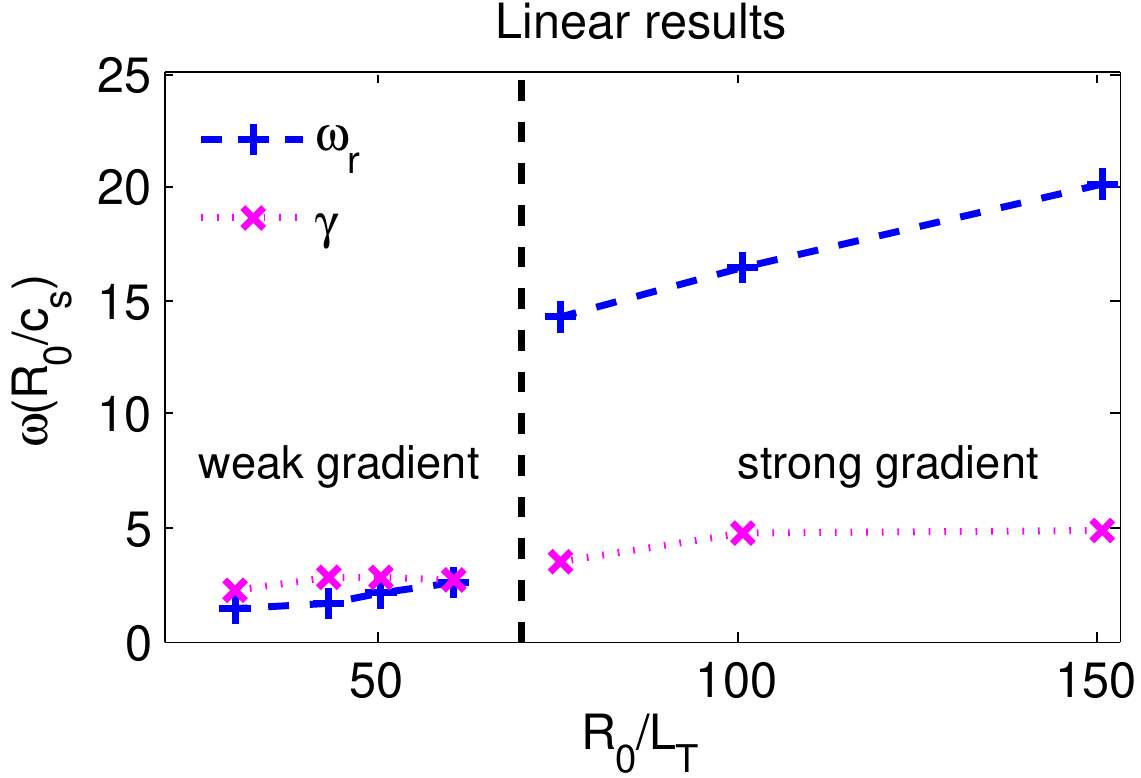}\\
\caption{Linear real frequency and growth rate for most unstable
mode under different temperature gradients using typical HL-2A
parameters with selected toroidal mode number $n=20$.
}\label{fig:scan_grad_lin}
\end{figure}

The preceding nonlinear results in Fig. \ref{fig:flux_vs_grad} and
Fig. \ref{fig:hsty_cmp_zf_lh} can be further understood by a random
walk model, i.e., the transport coefficients such as the particle
diffusivity  $D$, agree with $D\sim l_c^2/\tau_c$, where $l_c$ is
the correlation length and $\tau_c$ is the correlation time. By
linking $\sim l_c$ with characteristic perpendicular wavelength and
$\tau_c$ with the linear growth rate, a simplest model of this type
gives $D\sim(\gamma_k/k^2_\perp)\propto\gamma_k$. A rough estimate
for the linear growth rate gives $\gamma_k\propto\nabla T$, thus we
have $D\propto\nabla T$, which leads to the Taroni-Bohm model
\cite{Horton2012}. However, this model breaks down for large
gradients, as shown in Fig. \ref{fig:flux_vs_grad}, as the linear
growth rate $\gamma_k$ (inverse of the correlation time $\tau_c$)
does not increase much with the gradient {\color{red}{(see  Fig.
\ref{fig:scan_grad_lin})}}. On the other hand, the correlation
length $l_c$ should be reduced dramatically in the strong gradient
region, as implied by Fig. \ref{fig:hsty_cmp_scangrad_h}. Fig.
\ref{fig:phirz} shows two typical eddy size for the nonlinearly
strong gradient ($R_0/L_T=30$ ) and the nonlinearly weak gradient
($R_0/L_T=10$). The zonal flow is excluded in these simulations to
manifest the turbulence mode structure. We can see strong streamer
structures for the case with $R_0/L_T=10$, which shows elongated
radial eddies in Fig. \ref{fig:phirz}(a) and hence large radial
correlation length. The zonal flow can cut through and stretch these
streamer structures and therefore can effectively reduce the
transport. However, under stronger gradient $R_0/L_T=30$, the eddy
size becomes smaller and zonal flow may no longer cut through them,
which can minimize the regulation effect of the zonal flow, as shown
in Fig. \ref{fig:hsty_cmp_zf_lh}.

\begin{figure}
\centering
\includegraphics[width=8.5cm]{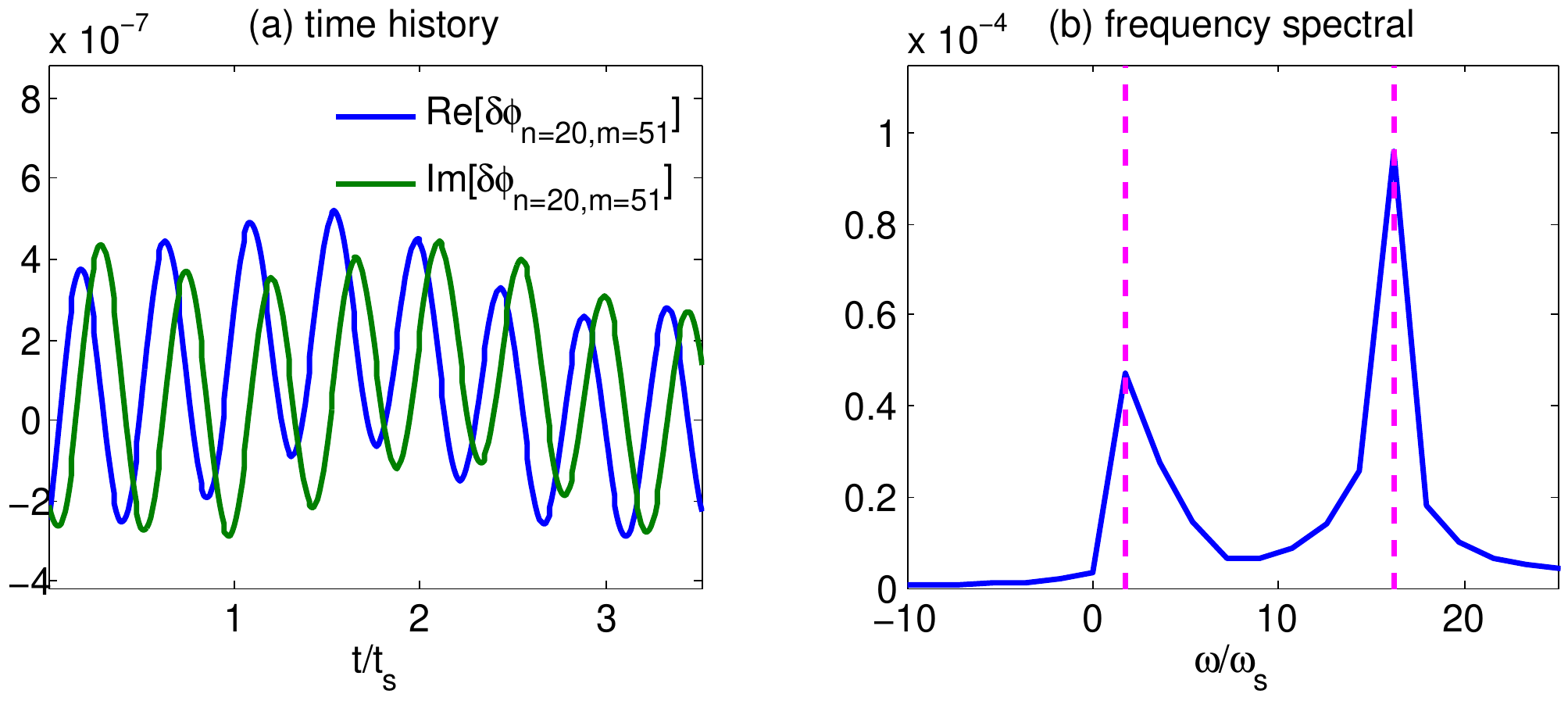}\\
\caption{The time history (a) and frequency spectral (b) for the
Fourier component ($n=20$, $m=51$) of electrostatic potential in the
linear simulation with $R_0/L_T=75$. }\label{fig:tem_lin_freq}
\end{figure}

The reverse trend of the transport coefficient in the strong
gradient region can be reasonably explained by the smaller eddy size
and hence smaller correlation length, which could possible be
induced by the unconventional mode structures of non-ground
eigenstates of micro-instabilities \cite{Xie2015b}. To examine the
physics mechanism for this reverse transport trend under strong
gradient, we thus further perform a linear simulation for the most
unstable mode with toroidal mode number $n=20$, since that the
discontinuous change in the nonlinear transport may be related to
the discontinuity in the linear eigenmode characteristics. As shown
by time history and spectrum analysis of selected unstable mode in
Fig. \ref{fig:tem_lin_freq} (for poloidal mode number $m=51$), two
distinct frequencies clearly coexist in electron diamagnetic
direction for the linear simulation. Suppose that the 3D (three
dimensional) mode structure of the electrostatic potential is
represented by the Fourier series
$\delta\phi(r,\theta,\zeta,t)=e^{in\zeta-i\omega t}\sum_m
\delta\phi_m(r)e^{-im\theta}$, where $\omega=\omega_r+i\gamma$ is
the mode frequency. We proceed to examine the mode frequency
variation under different gradients. As shown in Fig.
\ref{fig:scan_grad_lin}, there exists a clearly frequency jump from
low frequency ($\omega_r<3\omega_s$, $\omega_r<\gamma$) to high
frequency ($\omega_r>10\omega_s$, $\omega_r\gg\gamma$) branch, where
the normalized frequency $\omega_s=1/t_s\equiv c_s/R_0$ and
$c_s\equiv \sqrt{T_e/m_i}$. Around the critical jump gradient
($R_0/L_T\simeq70$), two branches of the eigenmode coexist at the
initial linear stage due to similar growth rates, as shown in Fig.
\ref{fig:tem_lin_freq}. The low frequency branch shows a
conventional ballooning structure localized at the outside mid-plane
(ground eigenstate), whereas the high frequency branch shows an
unconventional mode structure which can localize at almost arbitrary
poloidal positions or with multiple peaks (non-ground eigenstate)
\cite{Xie2015,Xie2015b}. For the weak gradient (L-mode), the most
unstable mode is in the ground state. However, the most unstable
mode can jump to non-ground eigenstate under strong gradient
(H-mode). The unconventional mode structure can significantly reduce
the effective correlation length and thus the turbulent transport.
This viewpoint can provide a mechanism to understand the previous
nonlinear simulations in this paper. We note the critical gradient
for the frequency jump is $R_0/L_T\simeq70$, which is around the
experimental L-H transition gradient $R_0/L_T\simeq40-120$
\cite{Kong2017,Ryter2016}. The low and high frequencies from the
simulation also quantitatively agree with the characteristic
frequencies of the electrostatic turbulence for typical HL-2A L-mode
and H-mode, i.e., $\sim20kHz$ and $\sim80kHz$, respectively
\cite{Xie2015}. The detailed linear results, especially the
unconventional mode structures and eigenstate jump, have been
reported in Refs. \cite{Xie2015,Xie2015b}, where the
mirco-instability is identified as trapped electron modes (TEMs)
\cite{Coppi1974,Catto1978}. It has also been recently reported that
the turbulence (usually also TEM) jump from low frequency to high
frequency during the L-H transition for experiments such as DIII-D
\cite{Barada2015} or EAST \cite{Xu2012}. Other simulations also
discussed the possible important roles of resistive ballooning mode
\cite{Bourdelle2015} or the large diamagnetic frequency
\cite{Scott2010}, but they did not show a clear sudden change of
transport characteristics as the present work does.

\begin{figure}
\centering
\includegraphics[width=8.5cm]{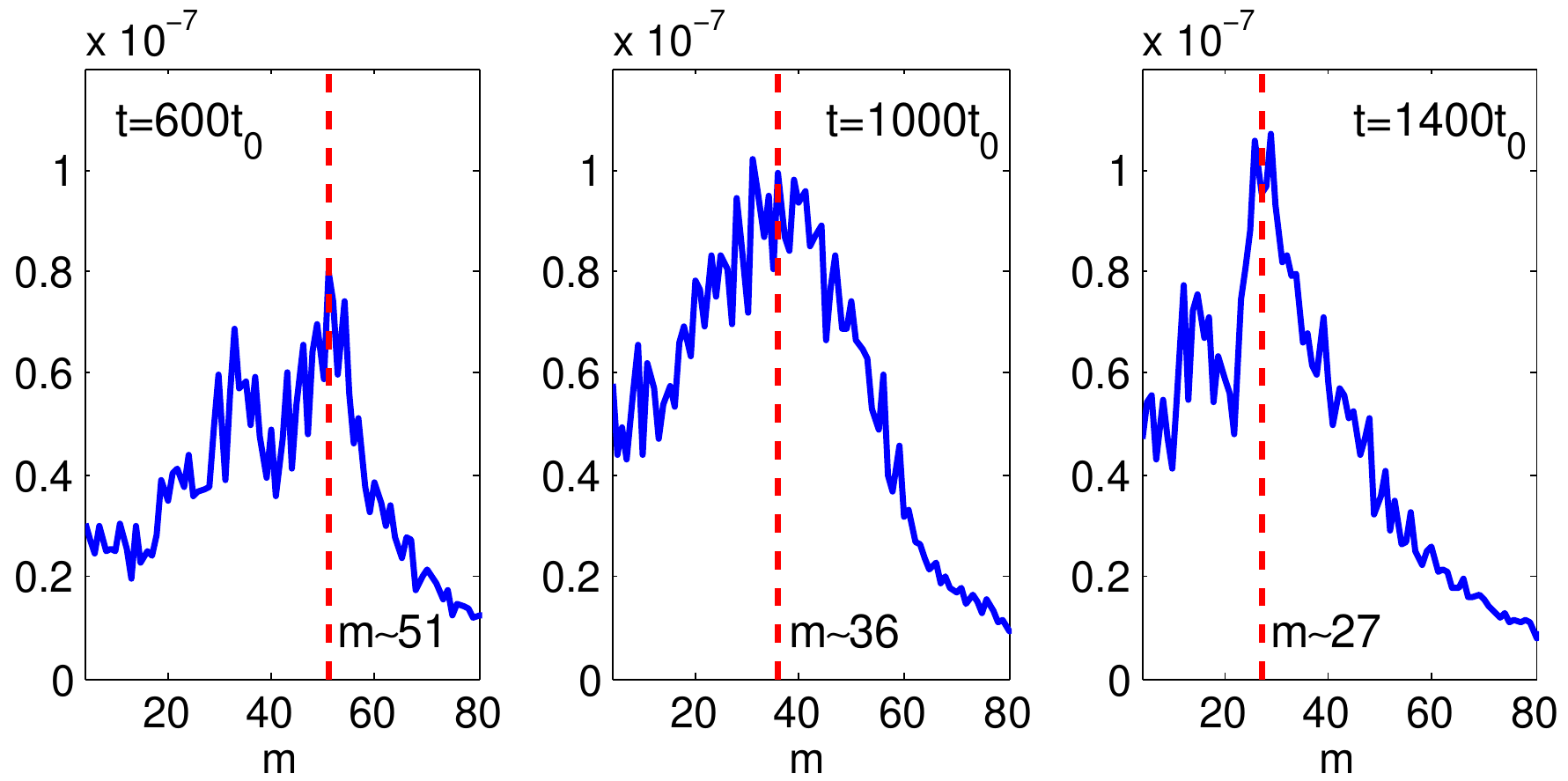}\\
\caption{The poloidal spectral cascade from high to low mode number
during nonlinear saturation.}\label{fig:snap_spectral}
\end{figure}

So far we have discovered the coexistence of both linear and
nonlinear critical gradients. However, the linear critical gradient
($R_0/L_T=75$) is larger than the nonlinear critical gradient
($R_0/L_T=25$). There could have two reasons for this difference.
The first one is shown by Fig. \ref{fig:scan_grad_lin}: the linear
growth rate ceases to grow before the linear discontinuity occurs.
The second reason is associated with the following inverse poloidal
spectral cascade in the nonlinear saturation of turbulence. We carry
out a nonlinear gyrokinetic simulation under strong gradient with a
typical HL-2A H-mode experimental value $R/L_T=150$. As shown in
Fig. \ref{fig:snap_spectral}, the peaking $m$ downshifts from a
larger number to a smaller number during the nonlinear saturation
process, which demonstrates a nonlinear inverse cascade in the
poloidal spectrum. The peaking $m$ at a later saturation stage
($t=1400t_0$, steady state, where $t_0$ is time step size) is
$m=10-40$, whose value is close to the experimental value $m=14-33$
\cite{Kong2017,Xie2015}.

In conclusion, via first principle gyrokinetic simulation we have
found a trend reversal in the transport coefficients and a
discontinuous change of the slope of the transport flux in the
strong-gradient regime of magnetic fusion plasmas, indicating that a
small increase of the heat flux can lead to a large increase of the
gradient, similar to a second order phase transition
\cite{Landau1937}. We also found that there exist both linear and
nonlinear critical gradients for the discontinuity of the transport
characteristics. In the linear simulation, with increase of the edge
gradient, the most unstable mode jumps from the ground eigenstate to
another eigenstate. The unconventional mode structure associated
with the latter can effectively reduce the correlation length and
thus the transport coefficients. This result is confirmed by the
nonlinear simulation, which shows that the radial correlation length
is indeed reduced in the strong gradient regime, and a turning point
(the nonlinear critical gradient) indeed appears in the transport
coefficients. The reduction of the critical gradients and transport
coefficient can be crucial to the formation of the H-mode external
transport barrier (ETB) as well as the L-H transition. This result
therefore suggests a new pathway to the H-mode regime, namely
without the need of shear and/or zonal flows. In fact, experiments
have also raised doubt on the need of the latter for the L-H
transition \cite{Kobayashi2013,Zweben2010}, and the corresponding
fluid models for the L-H transition are also not fully convincing
\cite{Scott2010}. Finally, we note that gyrokinetic simulations can
provide quantitative outputs for closer comparison with experimental
results. In fact, the critical gradient, characteristic frequency
and poloidal mode number from our nonlinear simulation are
consistent with the HL-2A experiments. Moreover, for further
resolving the mystery of the L-H transition, other effects such as
flow shear, electromagnetic perturbations, self-consistent evolution
of the plasma profiles, etc. can also be included in the gyrokinetic
simulation.

{\it Acknowledgments} HSX would like to thank D. F. Kong, G. S. Xu,
and K. K. Barada for providing the experimental information form the
HL-2A, EAST, and DIIID tokamaks. We would like to thank L. Chen, G.
Y. Fu and M. Y. Yu of ZJU, P. H. Diamond of UCSD and X. Q. Xu of
LLNL for useful discussions. This work is supported by the National
Magnetic Confinement Fusion Energy Research Program under Grant Nos.
2015GB110000, 2013GB111000, the China NSFC under Grant Nos.
11575158, the Recruitment Program of Global Youth Experts, and the
US DOE SciDac GSEP Center.

\end{document}